\documentclass[prl,twocolumn,aps,superscriptaddress,showpacs,nofootinbib]{revtex4-1}
\usepackage{amsfonts}
\usepackage{amsmath}
\usepackage{amsthm}
\usepackage{amscd}
\usepackage{amssymb}
\usepackage{subfigure}
\usepackage{amsxtra}
\usepackage{gensymb}
\usepackage{bm}           
\usepackage{bbm}
\usepackage{graphicx}
\usepackage{epstopdf}
\usepackage{color}
\usepackage{times}
\usepackage{mdframed}
\usepackage{textpos}
\usepackage[colorlinks=true,citecolor=blue,urlcolor=black]{hyperref}

\def\bi#1\ei {\begin{itemize}#1\end{itemize}}
\def\bn#1\en {\begin{enumerate}#1\end{enumerate}}
\def\bea#1\eea {\begin{align}#1\end{align}}
\def\bean#1\eean {\begin{align*}#1\end{align*}}
\def\ben#1\een {\begin{equation*}#1\end{equation*}}
\def\be#1\ee {\begin{equation}#1\end{equation}}
\def\bes#1\ees {\begin{equation}\begin{split}#1\end{split}\end{equation}}
\def\bear#1\eear {\begin{eqnarray}#1\end{eqnarray}}
\def\bear#1\eear {\begin{eqnarray*}#1\end{eqnarray*}}

\newcommand{\beq}{\begin{equation}}
\newcommand{\eeq}{\end{equation}}

\begin{document}

\title{\bf Improved Spatial Resolution Achieved by Chromatic Intensity Interferometry}
\begin{textblock*}{5cm}(14.5cm,-1.5cm)
  \fbox{\footnotesize MIT-CTP-5111}
\end{textblock*}

\author{Lu-Chuan Liu}
\affiliation{Hefei National Laboratory for Physical Sciences at the Microscale and Department of Modern Physics, University of Science and Technology of China, Hefei 230026, China}
\affiliation{Shanghai Branch, CAS Center for Excellence in Quantum Information and Quantum Physics, University of Science and Technology of China, Shanghai 201315, China}
\affiliation{Shanghai Research Center for Quantum Sciences, Shanghai 201315, China}

\author{Luo-Yuan Qu}
\affiliation{Hefei National Laboratory for Physical Sciences at the Microscale and Department of Modern Physics, University of Science and Technology of China, Hefei 230026, China}
\affiliation{Shanghai Branch, CAS Center for Excellence in Quantum Information and Quantum Physics, University of Science and Technology of China, Shanghai 201315, China}
\affiliation{Shanghai Research Center for Quantum Sciences, Shanghai 201315, China}
\affiliation{Jinan Institute of Quantum Technology, Jinan, 250101, P.~R.~China}

\author{Cheng Wu}
\affiliation{Hefei National Laboratory for Physical Sciences at the Microscale and Department of Modern Physics, University of Science and Technology of China, Hefei 230026, China}
\affiliation{Shanghai Branch, CAS Center for Excellence in Quantum Information and Quantum Physics, University of Science and Technology of China, Shanghai 201315, China}
\affiliation{Shanghai Research Center for Quantum Sciences, Shanghai 201315, China}

\author{Jordan Cotler}

\affiliation{Society of Fellows, Harvard University, Cambridge, MA 02138 USA}

\author{Fei Ma}
\affiliation{Hefei National Laboratory for Physical Sciences at the Microscale and Department of Modern Physics, University of Science and Technology of China, Hefei 230026, China}
\affiliation{Shanghai Branch, CAS Center for Excellence in Quantum Information and Quantum Physics, University of Science and Technology of China, Shanghai 201315, China}
\affiliation{Shanghai Research Center for Quantum Sciences, Shanghai 201315, China}
\affiliation{Jinan Institute of Quantum Technology, Jinan, 250101, P.~R.~China}

\author{Ming-Yang Zheng}
\author{Xiu-Ping Xie}
\affiliation{Jinan Institute of Quantum Technology, Jinan, 250101, P.~R.~China}

\author{Yu-Ao Chen}
\affiliation{Hefei National Laboratory for Physical Sciences at the Microscale and Department of Modern Physics, University of Science and Technology of China, Hefei 230026, China}
\affiliation{Shanghai Branch, CAS Center for Excellence in Quantum Information and Quantum Physics, University of Science and Technology of China, Shanghai 201315, China}
\affiliation{Shanghai Research Center for Quantum Sciences, Shanghai 201315, China}

\author{Qiang Zhang}
\affiliation{Hefei National Laboratory for Physical Sciences at the Microscale and Department of Modern Physics, University of Science and Technology of China, Hefei 230026, China}
\affiliation{Shanghai Branch, CAS Center for Excellence in Quantum Information and Quantum Physics, University of Science and Technology of China, Shanghai 201315, China}
\affiliation{Shanghai Research Center for Quantum Sciences, Shanghai 201315, China}
\affiliation{Jinan Institute of Quantum Technology, Jinan, 250101, P.~R.~China}

\author{Frank Wilczek}
\affiliation{Center for Theoretical Physics, MIT, Cambridge, MA 02139 USA}
\affiliation{T. D. Lee Institute, Shanghai Jiao Tong University, Shanghai, 200240, P.~R.~China}
\affiliation{Wilczek Quantum Center, School of Physics and Astronomy, Shanghai Jiao Tong University, Shanghai, 200240, P.~R.~China}
\affiliation{Department of Physics, Stockholm University, Stockholm SE-106 91 Sweden}
\affiliation{Department of Physics and Origins Project, Arizona State University, Tempe, AZ 25287 USA}

\author{Jian-Wei Pan}
\affiliation{Hefei National Laboratory for Physical Sciences at the Microscale and Department of Modern Physics, University of Science and Technology of China, Hefei 230026, China}
\affiliation{Shanghai Branch, CAS Center for Excellence in Quantum Information and Quantum Physics, University of Science and Technology of China, Shanghai 201315, China}
\affiliation{Shanghai Research Center for Quantum Sciences, Shanghai 201315, China}


\begin{abstract}
Interferometers are widely used in imaging technologies to achieve enhanced spatial resolution, but require that the incoming photons be indistinguishable. In previous work, we built and analyzed color erasure detectors which expand the scope of intensity interferometry to accommodate sources of different colors.  Here we experimentally demonstrate how color erasure detectors can achieve improved spatial resolution in an imaging task, well beyond the diffraction limit. Utilizing two 10.9 mm-aperture telescopes and a 0.8 m baseline, we measure the distance between a 1063.6 nm source and a 1064.4 nm source separated by 4.2 mm at a distance of 1.43 km, which surpasses the diffraction limit of a single telescope by about 40 times.  Moreover, chromatic intensity interferometry allows us to recover the phase of the Fourier transform of the imaged objects -- a quantity that is, in the presence of modest noise, inaccessible to conventional intensity interferometry. \end{abstract}

\maketitle

Since Hanbury Brown and Twiss (HBT) first proposed an ingenious method to exploit second-order interference~\cite{brown1956test,brown1957interferometry,brown1974intensity}, there has been a revolution in high-resolution imaging.
By correlating signals collected by separated detectors, the HBT intensity interferometer can surpass the resolving power of individual detectors in several diverse circumstances. HBT interferometry has been applied in many fields ranging from astronomy to nuclear and elementary particle physics. For example, several large HBT interferometers have demonstrated their superiority in high-resolution imaging of astronomical targets (most notably, the COAST~\cite{baldwin1996first}, the CHARA~\cite{ten2005first} and VLT interferometers~\cite{petrov2007amber}).  Related methods have also been used successfully to probe nuclear collisions~\cite{Baym1997ce}, to measure the quantum state of Bose-Einstein condensates~\cite{folling2005spatial,schellekens2005hanbury,ottl2005correlations,polkovnikov2006interference,cayla2020hanbury}, and to identify complex quantum phases in ultracold bosonic and fermionic atom systems~\cite{Henny1999,jeltes2007comparison,dall2013ideal}.

A drawback of conventional interferometric methods is that they only allow interference between photons of the same wavelength.  The information encoded in the correlations between photons of different wavelengths has attracted increasing attention in recent years~\cite{cotler2015entanglement, kobayashi2016frequency, kobayashi2017mach, lu2018electro, lu2018quantum, kues2019quantum}. The color erasure detector is a fundamental tool to realize chromatic interferometry~\cite{cotler2016entanglement, qu2019color, qu2020chromatic}, and recover the hidden information. Unlike previous experiments in chromatic interferometry, which implemented wavelength conversion either at the light sources themselves, or nearby~\cite{takesue2008erasing, raymer2010interference, de2012quantum, kobayashi2016frequency, kobayashi2017mach, lu2018quantum}, our color erasure detectors operate only on photons at the final detection stage. This feature of color erasure detectors allows them to interface smoothly with conventional intensity interferometry~\cite{brown1956test,twiss1957question,baym1998physics}.

Major advantages of the optical intensity interferometer are its relative insensitivity to atmospheric turbulence and the fact that it does not require sub-wavelength precision in the optical components and delay lines~\cite{brown1974intensity}.  According to the van Cittert-Zernike theorem~\cite{van1934wahrscheinliche,zernike1938concept}, however, traditional intensity interferometry only obtains the squared-magnitude of the Fourier transform of the radiance distribution of an imaged object.  This loss of phase information poses a severe difficulty in the reconstruction of images. Chromatic intensity interferometry not only achieves interference between sources of different wavelengths, but also obtains the phase of the Fourier transform.

In this paper we demonstrate, theoretically and experimentally, that chromatic intensity interferometry can improve spatial resolution in imaging.  In our experiment, we spatially resolve a 1064.4 nm source and a 1063.6 nm source separated by 4.2 mm at a distance of 1.43 km, by measuring the second order correlation of the signal light collected via color erasure detectors.  

We determine the distance between the light sources as follows.  Suppose our interferometer consists of two detectors, namely telescopes $T_A$ and $T_B$ positioned at $\mathbf{r}_A$ and $\mathbf{r}_B$, respectively.  We want to resolve two distant point sources emitting at different wavelengths that are spatially close to one another.  If the two sources, labeled $S_1$ and $S_2$, are positioned at $\mathbf{r}_1,\mathbf{r}_2$ and emit photons with wavelengths $\lambda_1,\lambda_2$, respectively, then the phase $\phi_s$ of the intensity interference due to the spatial optical path has the form~\cite{cotler2016entanglement}
\beq
\phi_s=\frac{2\pi}{\lambda_1}(|\mathbf{r}_1-\mathbf{r}_B|-|\mathbf{r}_1-\mathbf{r}_A|)-\frac{2\pi}{\lambda_2}(|\mathbf{r}_2-\mathbf{r}_B|-|\mathbf{r}_2-\mathbf{r}_A|)\,.
\eeq

\begin{figure}[t!]
\centering
\resizebox{9cm}{!}{\includegraphics{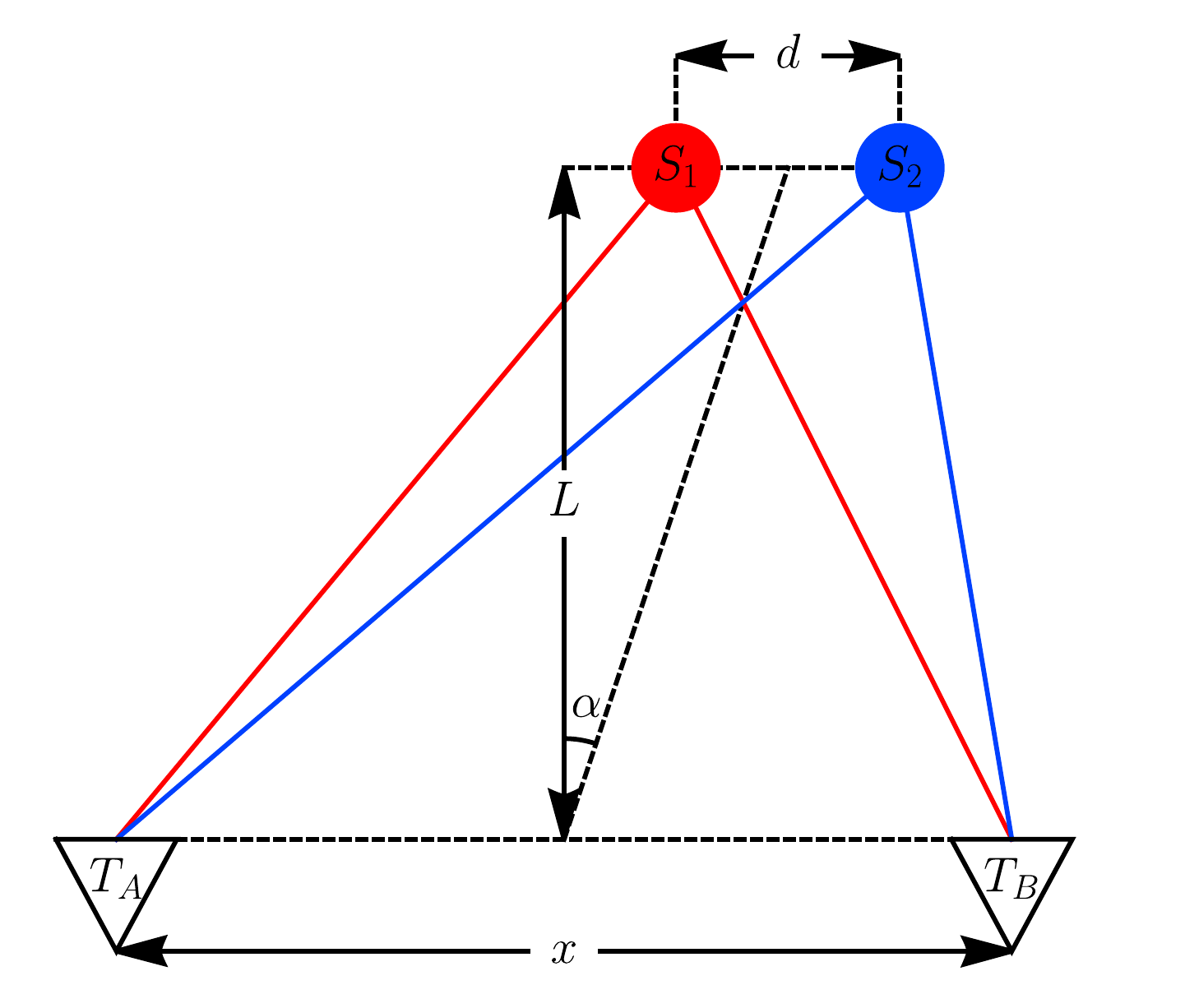}}
\caption{\textbf{Geometry of the intensity interferometer.} Consider a coordinate system with the interferometer baseline and its perpendicular bisector as the axes.  Then the telescopes $T_A$ and $T_B$ are positioned at $\mathbf{r}_A=(-x/2,0)$ and $\mathbf{r}_B=(x/2,0)$, respectively, and the sources $S_1$ and $S_2$ are positioned at $(L\alpha-d/2,L)$ and $(L\alpha+d/2,L)$ in the small angle approximation, respectively.}
\label{fig1}
\end{figure}
Adjusting the direction of the interferometer baseline so that the target falls on its perpendicular bisector, we arrive at the geometry shown in Fig.~\ref{fig1}, where $x$ is the distance between the two telescopes, $L$ is the distance from the target to the baseline, $d$ is the projected distance of the two sources onto the baseline, and $\alpha$ is the angle between the perpendicular bisector of the baseline and the midpoint of the two sources.  We work in a regime where the parameters satisfy the condition $x/L,\,d/L,\,\alpha\ll1$.  In this regime, we have
\beq
\label{E:phiS}
\phi_s=\frac{2\pi x}{\lambda_h}\left(\theta+\frac{\alpha\Delta\lambda}{\lambda_a}\right),
\eeq
where $\lambda_a,\lambda_h$ and $\Delta\lambda$ are given by
\beq
\begin{aligned}
\lambda_a\!=\!\frac{\lambda_1\!+\!\lambda_2}{2}\,,\quad \lambda_h\!=\!\frac{2\lambda_1 \lambda_2}{\lambda_1\!+\!\lambda_2}\,,\quad \Delta\lambda\!=\!\lambda_1\!-\!\lambda_2\,, \quad \theta\!=\!\frac{d}{L}\,.
\end{aligned}
\eeq

In practice, $\phi_s$ can be extracted by the second-order correlation function $g^{(2)}$ between the two color erasure detectors. If $S_1$ and $S_2$ are coherent sources and the delay $\tau$ applied to the signal of $T_B$ is much smaller than the coherence time of color-erased photons, a theoretical formula for $g^{(2)}(\tau)$ is~\cite{qu2020chromatic}
\beq
\label{E:g2theory}
g^{(2)}(\tau)=1+\frac{\epsilon}{2}\cos(2\pi(f_3^{(1)}-f_3^{(2)})\tau+\phi_{c})\,,
\eeq
where $\epsilon$ is the visibility, $f_3^{(1)}$ and $f_3^{(2)}$ are the frequencies of color-erased photons from source $S_1$ and $S_2$ respectively, and $\phi_{c}$ is the total phase of the intensity interference.  Here
\beq
\label{E:phiC}
\phi_{c}=\phi_{s}+\phi_{f}+\phi_{n}
\eeq
is the sum of the spatial phase $\phi_s$ defined previously, the inherent phase $\phi_f$ due to all the optical fibers that carry the pump light or signal light, and the noise phase $\phi_n$ caused by various factors including atmospheric disturbance and fiber deformation.

We remark that $\epsilon$ is itself a stochastic variable subject to complicated time-dependent drift.  In the standard HBT setting where $f_3^{(1)} = f_{3}^{(2)}$, Eqn.~\eqref{E:g2theory} becomes $1 + \frac{\epsilon}{2} \, \cos(\phi_c)$, and it is hard to estimate $\phi_c$ since the time-dependence of $\epsilon$ is difficult to characterize.  However, when $f_3^{(1)} \not= f_{3}^{(2)}$ in the color erasure setting, we can readily extract $\phi_c$ by examining the dependence of Eqn.~\eqref{E:g2theory} on $\tau$, crucially even when the time-dependence of $\epsilon$ is complicated.

In the color erasure setting, we can determine $\phi_c$ as a function of $x$ by measuring $g^{(2)}(\tau)$ for different $x$, and extracting $\phi_c$ using Eqn.~\eqref{E:g2theory}. We can regard $\phi_f$ as a constant and $\phi_n$ a random variable with mean $0$. If we ignore $\phi_n$ and substitute Eqn.~\eqref{E:phiS} into Eqn.~\eqref{E:phiC}, we can find that $\phi_c$ is a linear function of $x$, and so the slope $m = \frac{\partial \phi_c}{\partial x}$ can be written as
\beq
m=\frac{\partial\phi_c}{\partial x}=\frac{\partial\phi_s}{\partial x}=\frac{2\pi}{\lambda_h}\left(\theta+\frac{\alpha\Delta\lambda}{\lambda_a}\right)\,.
\eeq
Accordingly, $\theta$ has the form
\beq
\theta=\frac{m\lambda_h}{2\pi}-\frac{\alpha\Delta\lambda}{\lambda_a}\,.
\eeq

For our purposes, we will take $\alpha=0$ and include an angle adjustment uncertainty $\sigma_{\alpha}$.  For $m$, there is also an uncertainty $\sigma_m$ due to the stochastic phase $\phi_n$ defined previously. Then $\theta$ and its uncertainty $\sigma_{\theta}$ can be calculated as
\beq
\begin{aligned}
\label{E:theta}
\theta=\frac{m\lambda_h}{2\pi}\,,\quad \sigma_{\theta}=\sqrt{\left(\frac{\lambda_h}{2\pi}\right)^2\sigma_m^2+\left(\frac{\Delta\lambda}{\lambda_a}\right)^2\sigma_{\alpha}^2}\,\,.
\end{aligned}
\eeq

\begin{figure*}
\centering
\includegraphics[width=1\linewidth]{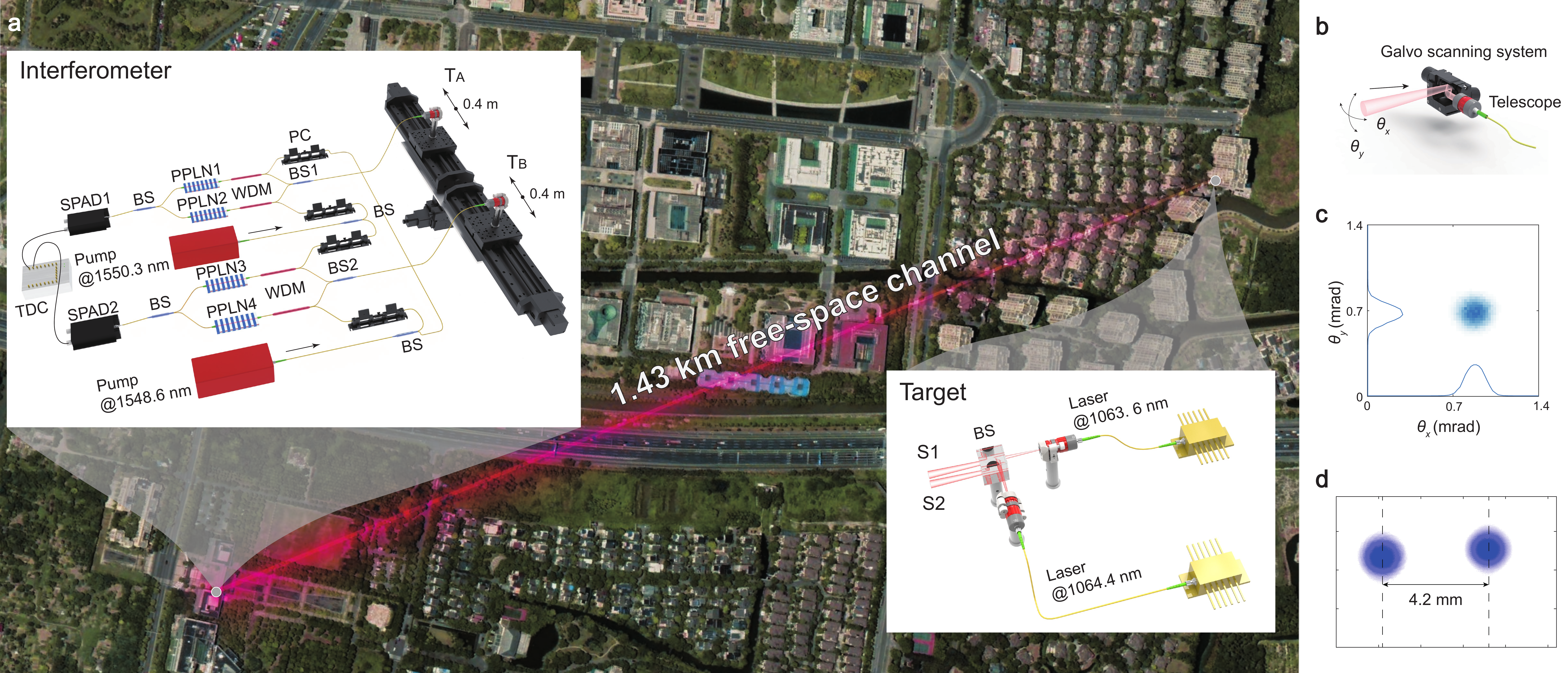}
\caption{\textbf{Scheme of the chromatic intensity interferometer.} (a) In the `Target' panel, two sources are coupled to free space by collimators. The 1063.6 nm light passes through the BS and the 1064.4 nm light is reflected, forming two point sources separated by 4.2 mm. The interferometer is 1.43 km away from the target. Two telescopes $T_{A}$ and $T_{B}$ move with a translation stage to change the baseline of the interferometer. Photons of different wavelengths received by $T_{A}$ ($T_{B}$) are divided by BS1 (BS2), and coupled into PPLN1 and PPLN2 (PPLN3 and PPLN4). PPLN1 (PPLN4) is pumped by a 1548.6 nm laser, and PPLN2 (PPLN3) is pumped by a 1550.3 nm laser. The polarization of the pump is controlled by PC. After SFG (sum-frequency generation), the upconverted 630.8 nm photons in PPLN1 and PPLN2 (PPLN3 and PPLN4) are combined by BS and guided to SPAD1 (SPAD2). The arrival time of the detected photons is recorded by TDC. By calculating the second order correlation of the signal recorded by TDC  with different baselines, we spatially resolve the two sources separated by 4.2 mm at the target. Some abbreviations are: beamsplitter (BS), polarization controller (PC), wavelength division multiplexing (WDM), periodically poled lithium niobate waveguide (PPLN), and Time-Digital Converter (TDC). (b) We depict the 2-axis galvo scanning system and the 10.9-mm telescope; this acts as a conventional (non-color erasure) telescope that we can benchmark against our color erasure intesnity interferometry.  This telescope collects photons from the same target at different $x$ and $y$ angles $\theta_x$ and $\theta_y$, controlled by the scanning system. Photons are captured with the same exposure time per sample.  (c) Intensity measurements using the non-color erasure telescopes with 64$\times$64 pixels cannot distinguish the distance between two light sources in the target. (d) The two sources are measured by a beam analyzer in close proximity to the sources, in order to calibrate the distance between their centers.  Using color erasure interferometry we can measure this distance from afar, as indicated below.}
\label{fig2}
\end{figure*}

Next we turn from our theoretical setup to an experimental demonstration of the resolution capabilities of chromatic intensity interferometry. As shown in Fig.~\ref{fig2}(a) and Fig.~\ref{fig2}(d), in a building $L=1.43$~km away from our laboratory, a $\lambda_1=1063.6$~nm transmission light and a $\lambda_2=1064.4$~nm reflected light form two sources separated horizontally by $d=4.2$ mm.  Each source has two emanating beams with a small relative angle; the beams are created by a beamsplitter.  We can see in Fig.~\ref{fig2}(b) and Fig.~\ref{fig2}(c) that a single 10.9~mm-aperture telescope nearby $T_1$ and $T_2$ cannot spatially resolve the two sources. The diffraction limit of a single 10.9~mm-aperture telescope is $1.9\times 10^{-4}$~rad when $\lambda=1064$~nm, which means it can only resolve sources separated by more than 0.17 m at a distance of 1.43 km. We utilize chromatic intensity interferometry shown in Fig.~\ref{fig2}(a) to resolve the sources. In our laboratory, two 10.9~mm-aperture telescopes are installed on two $0.4$~m-long translation stages, which move symmetrically to change $x$ from $0.16$~m to $0.96$~m. We place this system on a rotator to adjust the direction of the baseline to satisfy the $\alpha=0$ condition. Photons received by each telescope are guided to color erasure detectors of the same design as in~\cite{qu2020chromatic}. In a pair of parallel PPLN waveguides pumped by $1550.3$~nm and $1548.6$~nm lasers respectively, the received photons are respectively upconverted into indistinguishable (i.e,~$f_3^{(1)}\approx f_3^{(2)}$) $630.8$~nm photons via sum-frequency generation (SFG). A time-to-digital converter (TDC) is used to record the arrival time of these photons at two silicon single-photon avalanche diodes (SPAD), from which $g^{(2)}(\tau)$ can be calculated.

\begin{figure*}
\centering
\includegraphics[width=\linewidth]{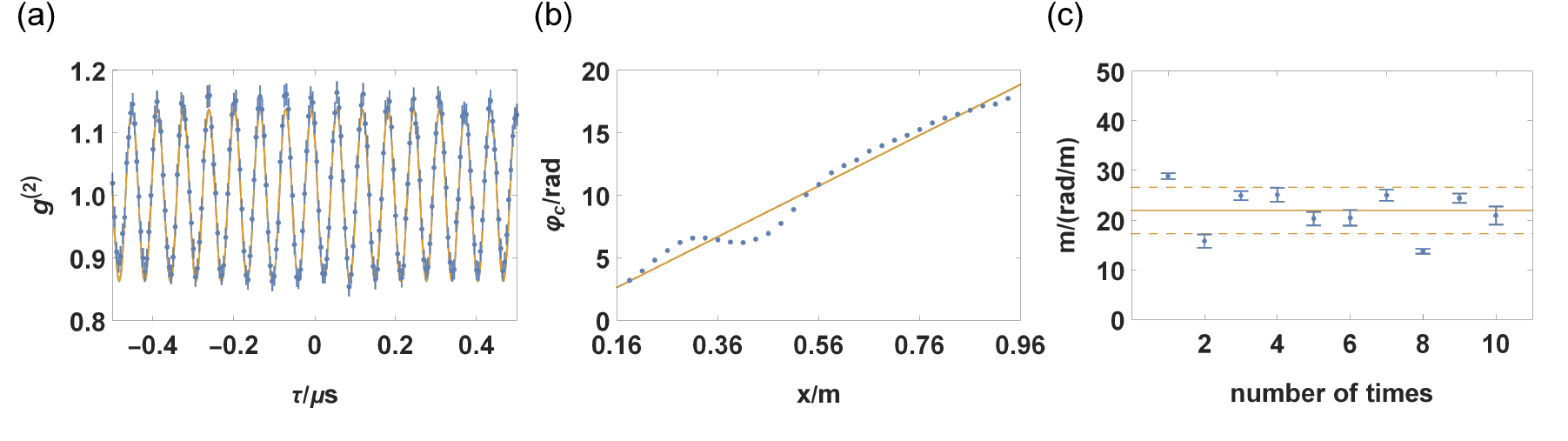}
\caption{\textbf{Experimental data for the chromatic intensity interferometer.} (a) A plot of the $g^{(2)}(\tau)$ measurement with optimized parameters. The orange curve is a least squares fit to the measured data (blue dots). (b) A graph of $\phi_c$ as a function of $x$ on the entire baseline. The orange line is a least squares fit to the measured data (blue dots). The slope of the fitted line provides a measurement of $m$. This set of data corresponds to the $5$th point in the next panel. (c) Shown are 10 measurements of $m$ and their distribution. The vertical position of the orange solid line is $\bar{m}$, and vertical positions of the two orange dashed lines are $\bar{m}+\sigma_m$ and $\bar{m}-\sigma_m$ respectively, where $\bar{m}$ and $\sigma_m$ are the average value and standard deviation calculated from the $10$ measurement samples (blue dots).}
\label{fig3}
\end{figure*}

Considering the time resolution capability of SPAD, when we fine-tune the frequencies of the pump lasers, we only need to make the value of $|f_3^{(1)}-f_3^{(2)}|$ reach the order of $10$~MHz. In addition, in order to make the interference visibility $\epsilon$ sufficiently large, the power and polarization of the pump lasers also need to be adjusted to control the SFG efficiency in each PPLN waveguide. Fig.~\ref{fig3}(a) shows the result of performing a $g^{(2)}(\tau)$ measurement using the optimized parameters $f_3^{(1)}-f_3^{(2)}=15.79\pm0.01$~MHz and $\epsilon=0.274\pm0.06$.  We let the two telescopes reciprocate symmetrically on the two translation stages, changing $x$ at a speed of $0.05$~m/s, and at the same time we continuously perform the $g^{(2)}(\tau)$ measurement. Every time the telescopes move from one end of the translation stages to the other, we obtain a data set of $\phi_c$ as a function of $x$ on the entire baseline and linearly fit the slope, as shown in Fig.~\ref{fig3}(b). We repeat this measurement $10$ times and obtain $10$ fitted lines. Since the value of $\phi_n$ drifts randomly in time, the slopes of these fitted lines have a distribution, as shown in Fig.~\ref{fig3}(c). Based on these samples, the unbiased estimates of the expected value of the slope $\bar{m}$ and the standard deviation $\sigma_m$ of the slope $m$ are
\beq
\begin{aligned}
\label{E:meas1}
\bar{m}&=22.0\,\text{ rad/m},
\\
\sigma_m&=4.6\,\text{ rad/m}.
\end{aligned}
\eeq
We also evaluate the standard deviation of the rotator angle as
\beq
\label{E:meas2}
\sigma_\alpha=1 \times 10^{-3}\,\text{ rad}.
\eeq
Substituting the results in~\eqref{E:meas1} and~\eqref{E:meas2} into Eqn.~\eqref{E:theta}, we finally get the expected value $\bar\theta$ and standard deviation $\sigma_\theta$ of $\theta$\,:
\beq
\begin{aligned}
\bar\theta&=3.7 \times 10^{-6}\,\text{ rad},
\\
\sigma_\theta&=1.1 \times 10^{-6}\,\text{ rad}.
\end{aligned}
\eeq

The actual value of $\theta$ is $d/L=2.93 \times 10^{-6}$~rad, which falls within one standard deviation of the experimental result. This result indicates that the spatial resolution of our interferometer has surpassed the diffraction limit of a $10$~mm-aperture telescope by about $40$ times. If we can further suppress the noise phase, the fitting of $\phi_c$ as a function of $x$ will become more accurate, enabling the interferometer to achieve a higher spatial resolution. Equivalently, a low signal-to-noise ratio will make $\phi_c$ more sensitive to small changes in $x$, so that the same level of spatial resolution can be achieved with a shorter baseline. Theoretically, if $\sigma_m=0$ is substituted into Eqn.~\eqref{E:theta}, the spatial resolution limit $\sigma_{\theta,\,\text{limit}}$ of the chromatic intensity interferometer in the infinite baseline limit has the form
\beq
\sigma_{\theta,\,\text{limit}}=\frac{\Delta\lambda}{\lambda_a}\,\sigma_\alpha\,,
\eeq
where we also assume that $\Delta \lambda$ is greater than the linewidth of the final detected photons.

In our work, we demonstrate the capability of chromatic intensity interferometer to achieve enhanced spatial resolution in a regime where existing imaging techniques fail. The main advantages of our scheme are that it expands the range of intensity interferometry to the multi-wavelength setting and gives us access to the phase of the Fourier transform of the imaged objects. To further improve the spatial resolution, chromatic intensity interferometry should be carefully designed to eliminate the internal phase noise, and $\sigma_\alpha$ should be pushed as low as possible.  Two resolve structures in two spatial dimensions, color erasure detectors will need to be combined with telescope arrays~\cite{nunez2012high, kieda2017stellar, abeysekara2020demonstration}.

\textbf{Acknowledgements}
This work was supported by the National Key R\&D Program of China (No.2018YFB0504300),the National Natural Science Foundation of China, the Chinese Academy of Sciences (CAS), Shanghai Municipal Science and Technology Major Project (Grant No.2019SHZDZX), and Anhui Initiative in Quantum Information Technologies. JC is supported by a Junior Fellowship from the Harvard Society of Fellows and the U.S. Department of Energy under grant Contract Number DE-SC0012567. FW's work is supported by the U.S. Department of Energy under grant Contract  Number DE-SC0012567, by the European Research Council under grant 742104, and by the Swedish Research Council under Contract No. 335-2014-7424. 

\bibliography{bibliography.bib}

\begin{thebibliography}{34}%
\makeatletter
\providecommand \@ifxundefined [1]{%
 \@ifx{#1\undefined}
}%
\providecommand \@ifnum [1]{%
 \ifnum #1\expandafter \@firstoftwo
 \else \expandafter \@secondoftwo
 \fi
}%
\providecommand \@ifx [1]{%
 \ifx #1\expandafter \@firstoftwo
 \else \expandafter \@secondoftwo
 \fi
}%
\providecommand \natexlab [1]{#1}%
\providecommand \enquote  [1]{``#1''}%
\providecommand \bibnamefont  [1]{#1}%
\providecommand \bibfnamefont [1]{#1}%
\providecommand \citenamefont [1]{#1}%
\providecommand \href@noop [0]{\@secondoftwo}%
\providecommand \href [0]{\begingroup \@sanitize@url \@href}%
\providecommand \@href[1]{\@@startlink{#1}\@@href}%
\providecommand \@@href[1]{\endgroup#1\@@endlink}%
\providecommand \@sanitize@url [0]{\catcode `\\12\catcode `\$12\catcode
  `\&12\catcode `\#12\catcode `\^12\catcode `\_12\catcode `\%12\relax}%
\providecommand \@@startlink[1]{}%
\providecommand \@@endlink[0]{}%
\providecommand \url  [0]{\begingroup\@sanitize@url \@url }%
\providecommand \@url [1]{\endgroup\@href {#1}{\urlprefix }}%
\providecommand \urlprefix  [0]{URL }%
\providecommand \Eprint [0]{\href }%
\providecommand \doibase [0]{http://dx.doi.org/}%
\providecommand \selectlanguage [0]{\@gobble}%
\providecommand \bibinfo  [0]{\@secondoftwo}%
\providecommand \bibfield  [0]{\@secondoftwo}%
\providecommand \translation [1]{[#1]}%
\providecommand \BibitemOpen [0]{}%
\providecommand \bibitemStop [0]{}%
\providecommand \bibitemNoStop [0]{.\EOS\space}%
\providecommand \EOS [0]{\spacefactor3000\relax}%
\providecommand \BibitemShut  [1]{\csname bibitem#1\endcsname}%
\let\auto@bib@innerbib\@empty
\bibitem [{\citenamefont {Brown}\ and\ \citenamefont
  {Twiss}(1956)}]{brown1956test}%
  \BibitemOpen
  \bibfield  {author} {\bibinfo {author} {\bibfnamefont {R.~H.}\ \bibnamefont
  {Brown}}\ and\ \bibinfo {author} {\bibfnamefont {R.}~\bibnamefont {Twiss}},\
  }\href@noop {} {\bibfield  {journal} {\bibinfo  {journal} {Nature}\ }\textbf
  {\bibinfo {volume} {178}},\ \bibinfo {pages} {1046} (\bibinfo {year}
  {1956})}\BibitemShut {NoStop}%
\bibitem [{\citenamefont {Brown}\ and\ \citenamefont
  {Twiss}(1957)}]{brown1957interferometry}%
  \BibitemOpen
  \bibfield  {author} {\bibinfo {author} {\bibfnamefont {R.~H.}\ \bibnamefont
  {Brown}}\ and\ \bibinfo {author} {\bibfnamefont {R.~Q.}\ \bibnamefont
  {Twiss}},\ }\href@noop {} {\bibfield  {journal} {\bibinfo  {journal}
  {Proceedings of the Royal Society of London. Series A. Mathematical and
  Physical Sciences}\ }\textbf {\bibinfo {volume} {242}},\ \bibinfo {pages}
  {300} (\bibinfo {year} {1957})}\BibitemShut {NoStop}%
\bibitem [{\citenamefont {Brown}(1974)}]{brown1974intensity}%
  \BibitemOpen
  \bibfield  {author} {\bibinfo {author} {\bibfnamefont {R.~H.}\ \bibnamefont
  {Brown}},\ }\href@noop {} {\emph {\bibinfo {title} {The intensity
  interferometer; its application to astronomy}}}\ (\bibinfo  {publisher}
  {London, Taylor \& Francis; New York, Halsted Press},\ \bibinfo {year}
  {1974})\BibitemShut {NoStop}%
\bibitem [{\citenamefont {Baldwin}\ \emph {et~al.}(1996)\citenamefont
  {Baldwin}, \citenamefont {Beckett}, \citenamefont {Boysen}, \citenamefont
  {Burns}, \citenamefont {Buscher}, \citenamefont {Cox}, \citenamefont
  {Haniff}, \citenamefont {Mackay}, \citenamefont {Nightingale}, \citenamefont
  {Rogers} \emph {et~al.}}]{baldwin1996first}%
  \BibitemOpen
  \bibfield  {author} {\bibinfo {author} {\bibfnamefont {J.}~\bibnamefont
  {Baldwin}}, \bibinfo {author} {\bibfnamefont {M.}~\bibnamefont {Beckett}},
  \bibinfo {author} {\bibfnamefont {R.}~\bibnamefont {Boysen}}, \bibinfo
  {author} {\bibfnamefont {D.}~\bibnamefont {Burns}}, \bibinfo {author}
  {\bibfnamefont {D.}~\bibnamefont {Buscher}}, \bibinfo {author} {\bibfnamefont
  {G.}~\bibnamefont {Cox}}, \bibinfo {author} {\bibfnamefont {C.}~\bibnamefont
  {Haniff}}, \bibinfo {author} {\bibfnamefont {C.}~\bibnamefont {Mackay}},
  \bibinfo {author} {\bibfnamefont {N.}~\bibnamefont {Nightingale}}, \bibinfo
  {author} {\bibfnamefont {J.}~\bibnamefont {Rogers}},  \emph {et~al.},\
  }\href@noop {} {\bibfield  {journal} {\bibinfo  {journal} {Astronomy and
  Astrophysics}\ }\textbf {\bibinfo {volume} {306}},\ \bibinfo {pages} {L13}
  (\bibinfo {year} {1996})}\BibitemShut {NoStop}%
\bibitem [{\citenamefont {Ten~Brummelaar}\ \emph {et~al.}(2005)\citenamefont
  {Ten~Brummelaar}, \citenamefont {McAlister}, \citenamefont {Ridgway},
  \citenamefont {Bagnuolo~Jr}, \citenamefont {Turner}, \citenamefont
  {Sturmann}, \citenamefont {Sturmann}, \citenamefont {Berger}, \citenamefont
  {Ogden}, \citenamefont {Cadman} \emph {et~al.}}]{ten2005first}%
  \BibitemOpen
  \bibfield  {author} {\bibinfo {author} {\bibfnamefont {T.}~\bibnamefont
  {Ten~Brummelaar}}, \bibinfo {author} {\bibfnamefont {H.}~\bibnamefont
  {McAlister}}, \bibinfo {author} {\bibfnamefont {S.}~\bibnamefont {Ridgway}},
  \bibinfo {author} {\bibfnamefont {W.}~\bibnamefont {Bagnuolo~Jr}}, \bibinfo
  {author} {\bibfnamefont {N.}~\bibnamefont {Turner}}, \bibinfo {author}
  {\bibfnamefont {L.}~\bibnamefont {Sturmann}}, \bibinfo {author}
  {\bibfnamefont {J.}~\bibnamefont {Sturmann}}, \bibinfo {author}
  {\bibfnamefont {D.}~\bibnamefont {Berger}}, \bibinfo {author} {\bibfnamefont
  {C.}~\bibnamefont {Ogden}}, \bibinfo {author} {\bibfnamefont
  {R.}~\bibnamefont {Cadman}},  \emph {et~al.},\ }\href@noop {} {\bibfield
  {journal} {\bibinfo  {journal} {The Astrophysical Journal}\ }\textbf
  {\bibinfo {volume} {628}},\ \bibinfo {pages} {453} (\bibinfo {year}
  {2005})}\BibitemShut {NoStop}%
\bibitem [{\citenamefont {Petrov}\ \emph {et~al.}(2007)\citenamefont {Petrov},
  \citenamefont {Malbet}, \citenamefont {Weigelt}, \citenamefont {Antonelli},
  \citenamefont {Beckmann}, \citenamefont {Bresson}, \citenamefont {Chelli},
  \citenamefont {Dugu{\'e}}, \citenamefont {Duvert}, \citenamefont {Gennari}
  \emph {et~al.}}]{petrov2007amber}%
  \BibitemOpen
  \bibfield  {author} {\bibinfo {author} {\bibfnamefont {R.}~\bibnamefont
  {Petrov}}, \bibinfo {author} {\bibfnamefont {F.}~\bibnamefont {Malbet}},
  \bibinfo {author} {\bibfnamefont {G.}~\bibnamefont {Weigelt}}, \bibinfo
  {author} {\bibfnamefont {P.}~\bibnamefont {Antonelli}}, \bibinfo {author}
  {\bibfnamefont {U.}~\bibnamefont {Beckmann}}, \bibinfo {author}
  {\bibfnamefont {Y.}~\bibnamefont {Bresson}}, \bibinfo {author} {\bibfnamefont
  {A.}~\bibnamefont {Chelli}}, \bibinfo {author} {\bibfnamefont
  {M.}~\bibnamefont {Dugu{\'e}}}, \bibinfo {author} {\bibfnamefont
  {G.}~\bibnamefont {Duvert}}, \bibinfo {author} {\bibfnamefont
  {S.}~\bibnamefont {Gennari}},  \emph {et~al.},\ }\href@noop {} {\bibfield
  {journal} {\bibinfo  {journal} {Astronomy \& Astrophysics}\ }\textbf
  {\bibinfo {volume} {464}},\ \bibinfo {pages} {1} (\bibinfo {year}
  {2007})}\BibitemShut {NoStop}%
\bibitem [{\citenamefont {Baym}(1998{\natexlab{a}})}]{Baym1997ce}%
  \BibitemOpen
  \bibfield  {author} {\bibinfo {author} {\bibfnamefont {G.}~\bibnamefont
  {Baym}},\ }\href@noop {} {\bibfield  {journal} {\bibinfo  {journal} {Acta
  Phys. Polon. B}\ }\textbf {\bibinfo {volume} {29}},\ \bibinfo {pages} {1839}
  (\bibinfo {year} {1998}{\natexlab{a}})},\ \Eprint
  {http://arxiv.org/abs/nucl-th/9804026} {nucl-th/9804026} \BibitemShut
  {NoStop}%
\bibitem [{\citenamefont {F{\"o}lling}\ \emph {et~al.}(2005)\citenamefont
  {F{\"o}lling}, \citenamefont {Gerbier}, \citenamefont {Widera}, \citenamefont
  {Mandel}, \citenamefont {Gericke},\ and\ \citenamefont
  {Bloch}}]{folling2005spatial}%
  \BibitemOpen
  \bibfield  {author} {\bibinfo {author} {\bibfnamefont {S.}~\bibnamefont
  {F{\"o}lling}}, \bibinfo {author} {\bibfnamefont {F.}~\bibnamefont
  {Gerbier}}, \bibinfo {author} {\bibfnamefont {A.}~\bibnamefont {Widera}},
  \bibinfo {author} {\bibfnamefont {O.}~\bibnamefont {Mandel}}, \bibinfo
  {author} {\bibfnamefont {T.}~\bibnamefont {Gericke}}, \ and\ \bibinfo
  {author} {\bibfnamefont {I.}~\bibnamefont {Bloch}},\ }\href@noop {}
  {\bibfield  {journal} {\bibinfo  {journal} {Nature}\ }\textbf {\bibinfo
  {volume} {434}},\ \bibinfo {pages} {481} (\bibinfo {year}
  {2005})}\BibitemShut {NoStop}%
\bibitem [{\citenamefont {Schellekens}\ \emph {et~al.}(2005)\citenamefont
  {Schellekens}, \citenamefont {Hoppeler}, \citenamefont {Perrin},
  \citenamefont {Gomes}, \citenamefont {Boiron}, \citenamefont {Aspect},\ and\
  \citenamefont {Westbrook}}]{schellekens2005hanbury}%
  \BibitemOpen
  \bibfield  {author} {\bibinfo {author} {\bibfnamefont {M.}~\bibnamefont
  {Schellekens}}, \bibinfo {author} {\bibfnamefont {R.}~\bibnamefont
  {Hoppeler}}, \bibinfo {author} {\bibfnamefont {A.}~\bibnamefont {Perrin}},
  \bibinfo {author} {\bibfnamefont {J.~V.}\ \bibnamefont {Gomes}}, \bibinfo
  {author} {\bibfnamefont {D.}~\bibnamefont {Boiron}}, \bibinfo {author}
  {\bibfnamefont {A.}~\bibnamefont {Aspect}}, \ and\ \bibinfo {author}
  {\bibfnamefont {C.~I.}\ \bibnamefont {Westbrook}},\ }\href@noop {} {\bibfield
   {journal} {\bibinfo  {journal} {Science}\ }\textbf {\bibinfo {volume}
  {310}},\ \bibinfo {pages} {648} (\bibinfo {year} {2005})}\BibitemShut
  {NoStop}%
\bibitem [{\citenamefont {{\"O}ttl}\ \emph {et~al.}(2005)\citenamefont
  {{\"O}ttl}, \citenamefont {Ritter}, \citenamefont {K{\"o}hl},\ and\
  \citenamefont {Esslinger}}]{ottl2005correlations}%
  \BibitemOpen
  \bibfield  {author} {\bibinfo {author} {\bibfnamefont {A.}~\bibnamefont
  {{\"O}ttl}}, \bibinfo {author} {\bibfnamefont {S.}~\bibnamefont {Ritter}},
  \bibinfo {author} {\bibfnamefont {M.}~\bibnamefont {K{\"o}hl}}, \ and\
  \bibinfo {author} {\bibfnamefont {T.}~\bibnamefont {Esslinger}},\ }\href@noop
  {} {\bibfield  {journal} {\bibinfo  {journal} {Physical Review Letters}\
  }\textbf {\bibinfo {volume} {95}},\ \bibinfo {pages} {090404} (\bibinfo
  {year} {2005})}\BibitemShut {NoStop}%
\bibitem [{\citenamefont {Polkovnikov}\ \emph {et~al.}(2006)\citenamefont
  {Polkovnikov}, \citenamefont {Altman},\ and\ \citenamefont
  {Demler}}]{polkovnikov2006interference}%
  \BibitemOpen
  \bibfield  {author} {\bibinfo {author} {\bibfnamefont {A.}~\bibnamefont
  {Polkovnikov}}, \bibinfo {author} {\bibfnamefont {E.}~\bibnamefont {Altman}},
  \ and\ \bibinfo {author} {\bibfnamefont {E.}~\bibnamefont {Demler}},\
  }\href@noop {} {\bibfield  {journal} {\bibinfo  {journal} {Proceedings of the
  National Academy of Sciences}\ }\textbf {\bibinfo {volume} {103}},\ \bibinfo
  {pages} {6125} (\bibinfo {year} {2006})}\BibitemShut {NoStop}%
\bibitem [{\citenamefont {Cayla}\ \emph {et~al.}(2020)\citenamefont {Cayla},
  \citenamefont {Butera}, \citenamefont {Carcy}, \citenamefont {Tenart},
  \citenamefont {Herc{\'e}}, \citenamefont {Mancini}, \citenamefont {Aspect},
  \citenamefont {Carusotto},\ and\ \citenamefont
  {Cl{\'e}ment}}]{cayla2020hanbury}%
  \BibitemOpen
  \bibfield  {author} {\bibinfo {author} {\bibfnamefont {H.}~\bibnamefont
  {Cayla}}, \bibinfo {author} {\bibfnamefont {S.}~\bibnamefont {Butera}},
  \bibinfo {author} {\bibfnamefont {C.}~\bibnamefont {Carcy}}, \bibinfo
  {author} {\bibfnamefont {A.}~\bibnamefont {Tenart}}, \bibinfo {author}
  {\bibfnamefont {G.}~\bibnamefont {Herc{\'e}}}, \bibinfo {author}
  {\bibfnamefont {M.}~\bibnamefont {Mancini}}, \bibinfo {author} {\bibfnamefont
  {A.}~\bibnamefont {Aspect}}, \bibinfo {author} {\bibfnamefont
  {I.}~\bibnamefont {Carusotto}}, \ and\ \bibinfo {author} {\bibfnamefont
  {D.}~\bibnamefont {Cl{\'e}ment}},\ }\href@noop {} {\bibfield  {journal}
  {\bibinfo  {journal} {Physical Review Letters}\ }\textbf {\bibinfo {volume}
  {125}},\ \bibinfo {pages} {165301} (\bibinfo {year} {2020})}\BibitemShut
  {NoStop}%
\bibitem [{\citenamefont {Henny}\ \emph {et~al.}(1999)\citenamefont {Henny},
  \citenamefont {Oberholzer}, \citenamefont {Strunk}, \citenamefont {Heinzel},
  \citenamefont {Ensslin}, \citenamefont {Holland},\ and\ \citenamefont
  {Sch{\"o}nenberger}}]{Henny1999}%
  \BibitemOpen
  \bibfield  {author} {\bibinfo {author} {\bibfnamefont {M.}~\bibnamefont
  {Henny}}, \bibinfo {author} {\bibfnamefont {S.}~\bibnamefont {Oberholzer}},
  \bibinfo {author} {\bibfnamefont {C.}~\bibnamefont {Strunk}}, \bibinfo
  {author} {\bibfnamefont {T.}~\bibnamefont {Heinzel}}, \bibinfo {author}
  {\bibfnamefont {K.}~\bibnamefont {Ensslin}}, \bibinfo {author} {\bibfnamefont
  {M.}~\bibnamefont {Holland}}, \ and\ \bibinfo {author} {\bibfnamefont
  {C.}~\bibnamefont {Sch{\"o}nenberger}},\ }\href {\doibase
  10.1126/science.284.5412.296} {\bibfield  {journal} {\bibinfo  {journal}
  {Science}\ }\textbf {\bibinfo {volume} {284}},\ \bibinfo {pages} {296}
  (\bibinfo {year} {1999})}\BibitemShut {NoStop}%
\bibitem [{\citenamefont {Jeltes}\ \emph {et~al.}(2007)\citenamefont {Jeltes},
  \citenamefont {McNamara}, \citenamefont {Hogervorst}, \citenamefont {Vassen},
  \citenamefont {Krachmalnicoff}, \citenamefont {Schellekens}, \citenamefont
  {Perrin}, \citenamefont {Chang}, \citenamefont {Boiron}, \citenamefont
  {Aspect} \emph {et~al.}}]{jeltes2007comparison}%
  \BibitemOpen
  \bibfield  {author} {\bibinfo {author} {\bibfnamefont {T.}~\bibnamefont
  {Jeltes}}, \bibinfo {author} {\bibfnamefont {J.~M.}\ \bibnamefont
  {McNamara}}, \bibinfo {author} {\bibfnamefont {W.}~\bibnamefont
  {Hogervorst}}, \bibinfo {author} {\bibfnamefont {W.}~\bibnamefont {Vassen}},
  \bibinfo {author} {\bibfnamefont {V.}~\bibnamefont {Krachmalnicoff}},
  \bibinfo {author} {\bibfnamefont {M.}~\bibnamefont {Schellekens}}, \bibinfo
  {author} {\bibfnamefont {A.}~\bibnamefont {Perrin}}, \bibinfo {author}
  {\bibfnamefont {H.}~\bibnamefont {Chang}}, \bibinfo {author} {\bibfnamefont
  {D.}~\bibnamefont {Boiron}}, \bibinfo {author} {\bibfnamefont
  {A.}~\bibnamefont {Aspect}},  \emph {et~al.},\ }\href@noop {} {\bibfield
  {journal} {\bibinfo  {journal} {Nature}\ }\textbf {\bibinfo {volume} {445}},\
  \bibinfo {pages} {402} (\bibinfo {year} {2007})}\BibitemShut {NoStop}%
\bibitem [{\citenamefont {Dall}\ \emph {et~al.}(2013)\citenamefont {Dall},
  \citenamefont {Manning}, \citenamefont {Hodgman}, \citenamefont {RuGway},
  \citenamefont {Kheruntsyan},\ and\ \citenamefont {Truscott}}]{dall2013ideal}%
  \BibitemOpen
  \bibfield  {author} {\bibinfo {author} {\bibfnamefont {R.}~\bibnamefont
  {Dall}}, \bibinfo {author} {\bibfnamefont {A.}~\bibnamefont {Manning}},
  \bibinfo {author} {\bibfnamefont {S.}~\bibnamefont {Hodgman}}, \bibinfo
  {author} {\bibfnamefont {W.}~\bibnamefont {RuGway}}, \bibinfo {author}
  {\bibfnamefont {K.~V.}\ \bibnamefont {Kheruntsyan}}, \ and\ \bibinfo {author}
  {\bibfnamefont {A.}~\bibnamefont {Truscott}},\ }\href@noop {} {\bibfield
  {journal} {\bibinfo  {journal} {Nature Physics}\ }\textbf {\bibinfo {volume}
  {9}},\ \bibinfo {pages} {341} (\bibinfo {year} {2013})}\BibitemShut {NoStop}%
\bibitem [{\citenamefont {Cotler}\ and\ \citenamefont
  {Wilczek}(2015)}]{cotler2015entanglement}%
  \BibitemOpen
  \bibfield  {author} {\bibinfo {author} {\bibfnamefont {J.}~\bibnamefont
  {Cotler}}\ and\ \bibinfo {author} {\bibfnamefont {F.}~\bibnamefont
  {Wilczek}},\ }\href@noop {} {\bibfield  {journal} {\bibinfo  {journal} {arXiv
  preprint arXiv:1502.02477}\ } (\bibinfo {year} {2015})}\BibitemShut {NoStop}%
\bibitem [{\citenamefont {Kobayashi}\ \emph {et~al.}(2016)\citenamefont
  {Kobayashi}, \citenamefont {Ikuta}, \citenamefont {Yasui}, \citenamefont
  {Miki}, \citenamefont {Yamashita}, \citenamefont {Terai}, \citenamefont
  {Yamamoto}, \citenamefont {Koashi},\ and\ \citenamefont
  {Imoto}}]{kobayashi2016frequency}%
  \BibitemOpen
  \bibfield  {author} {\bibinfo {author} {\bibfnamefont {T.}~\bibnamefont
  {Kobayashi}}, \bibinfo {author} {\bibfnamefont {R.}~\bibnamefont {Ikuta}},
  \bibinfo {author} {\bibfnamefont {S.}~\bibnamefont {Yasui}}, \bibinfo
  {author} {\bibfnamefont {S.}~\bibnamefont {Miki}}, \bibinfo {author}
  {\bibfnamefont {T.}~\bibnamefont {Yamashita}}, \bibinfo {author}
  {\bibfnamefont {H.}~\bibnamefont {Terai}}, \bibinfo {author} {\bibfnamefont
  {T.}~\bibnamefont {Yamamoto}}, \bibinfo {author} {\bibfnamefont
  {M.}~\bibnamefont {Koashi}}, \ and\ \bibinfo {author} {\bibfnamefont
  {N.}~\bibnamefont {Imoto}},\ }\href@noop {} {\bibfield  {journal} {\bibinfo
  {journal} {Nature photonics}\ }\textbf {\bibinfo {volume} {10}},\ \bibinfo
  {pages} {441} (\bibinfo {year} {2016})}\BibitemShut {NoStop}%
\bibitem [{\citenamefont {Kobayashi}\ \emph {et~al.}(2017)\citenamefont
  {Kobayashi}, \citenamefont {Yamazaki}, \citenamefont {Matsuki}, \citenamefont
  {Ikuta}, \citenamefont {Miki}, \citenamefont {Yamashita}, \citenamefont
  {Terai}, \citenamefont {Yamamoto}, \citenamefont {Koashi},\ and\
  \citenamefont {Imoto}}]{kobayashi2017mach}%
  \BibitemOpen
  \bibfield  {author} {\bibinfo {author} {\bibfnamefont {T.}~\bibnamefont
  {Kobayashi}}, \bibinfo {author} {\bibfnamefont {D.}~\bibnamefont {Yamazaki}},
  \bibinfo {author} {\bibfnamefont {K.}~\bibnamefont {Matsuki}}, \bibinfo
  {author} {\bibfnamefont {R.}~\bibnamefont {Ikuta}}, \bibinfo {author}
  {\bibfnamefont {S.}~\bibnamefont {Miki}}, \bibinfo {author} {\bibfnamefont
  {T.}~\bibnamefont {Yamashita}}, \bibinfo {author} {\bibfnamefont
  {H.}~\bibnamefont {Terai}}, \bibinfo {author} {\bibfnamefont
  {T.}~\bibnamefont {Yamamoto}}, \bibinfo {author} {\bibfnamefont
  {M.}~\bibnamefont {Koashi}}, \ and\ \bibinfo {author} {\bibfnamefont
  {N.}~\bibnamefont {Imoto}},\ }\href@noop {} {\bibfield  {journal} {\bibinfo
  {journal} {Optics Express}\ }\textbf {\bibinfo {volume} {25}},\ \bibinfo
  {pages} {12052} (\bibinfo {year} {2017})}\BibitemShut {NoStop}%
\bibitem [{\citenamefont {Lu}\ \emph {et~al.}(2018{\natexlab{a}})\citenamefont
  {Lu}, \citenamefont {Lukens}, \citenamefont {Peters}, \citenamefont {Odele},
  \citenamefont {Leaird}, \citenamefont {Weiner},\ and\ \citenamefont
  {Lougovski}}]{lu2018electro}%
  \BibitemOpen
  \bibfield  {author} {\bibinfo {author} {\bibfnamefont {H.-H.}\ \bibnamefont
  {Lu}}, \bibinfo {author} {\bibfnamefont {J.~M.}\ \bibnamefont {Lukens}},
  \bibinfo {author} {\bibfnamefont {N.~A.}\ \bibnamefont {Peters}}, \bibinfo
  {author} {\bibfnamefont {O.~D.}\ \bibnamefont {Odele}}, \bibinfo {author}
  {\bibfnamefont {D.~E.}\ \bibnamefont {Leaird}}, \bibinfo {author}
  {\bibfnamefont {A.~M.}\ \bibnamefont {Weiner}}, \ and\ \bibinfo {author}
  {\bibfnamefont {P.}~\bibnamefont {Lougovski}},\ }\href@noop {} {\bibfield
  {journal} {\bibinfo  {journal} {Physical Review Letters}\ }\textbf {\bibinfo
  {volume} {120}},\ \bibinfo {pages} {030502} (\bibinfo {year}
  {2018}{\natexlab{a}})}\BibitemShut {NoStop}%
\bibitem [{\citenamefont {Lu}\ \emph {et~al.}(2018{\natexlab{b}})\citenamefont
  {Lu}, \citenamefont {Lukens}, \citenamefont {Peters}, \citenamefont
  {Williams}, \citenamefont {Weiner},\ and\ \citenamefont
  {Lougovski}}]{lu2018quantum}%
  \BibitemOpen
  \bibfield  {author} {\bibinfo {author} {\bibfnamefont {H.-H.}\ \bibnamefont
  {Lu}}, \bibinfo {author} {\bibfnamefont {J.~M.}\ \bibnamefont {Lukens}},
  \bibinfo {author} {\bibfnamefont {N.~A.}\ \bibnamefont {Peters}}, \bibinfo
  {author} {\bibfnamefont {B.~P.}\ \bibnamefont {Williams}}, \bibinfo {author}
  {\bibfnamefont {A.~M.}\ \bibnamefont {Weiner}}, \ and\ \bibinfo {author}
  {\bibfnamefont {P.}~\bibnamefont {Lougovski}},\ }\href@noop {} {\bibfield
  {journal} {\bibinfo  {journal} {Optica}\ }\textbf {\bibinfo {volume} {5}},\
  \bibinfo {pages} {1455} (\bibinfo {year} {2018}{\natexlab{b}})}\BibitemShut
  {NoStop}%
\bibitem [{\citenamefont {Kues}\ \emph {et~al.}(2019)\citenamefont {Kues},
  \citenamefont {Reimer}, \citenamefont {Lukens}, \citenamefont {Munro},
  \citenamefont {Weiner}, \citenamefont {Moss},\ and\ \citenamefont
  {Morandotti}}]{kues2019quantum}%
  \BibitemOpen
  \bibfield  {author} {\bibinfo {author} {\bibfnamefont {M.}~\bibnamefont
  {Kues}}, \bibinfo {author} {\bibfnamefont {C.}~\bibnamefont {Reimer}},
  \bibinfo {author} {\bibfnamefont {J.~M.}\ \bibnamefont {Lukens}}, \bibinfo
  {author} {\bibfnamefont {W.~J.}\ \bibnamefont {Munro}}, \bibinfo {author}
  {\bibfnamefont {A.~M.}\ \bibnamefont {Weiner}}, \bibinfo {author}
  {\bibfnamefont {D.~J.}\ \bibnamefont {Moss}}, \ and\ \bibinfo {author}
  {\bibfnamefont {R.}~\bibnamefont {Morandotti}},\ }\href@noop {} {\bibfield
  {journal} {\bibinfo  {journal} {Nature Photonics}\ }\textbf {\bibinfo
  {volume} {13}},\ \bibinfo {pages} {170} (\bibinfo {year} {2019})}\BibitemShut
  {NoStop}%
\bibitem [{\citenamefont {Cotler}\ \emph {et~al.}(2016)\citenamefont {Cotler},
  \citenamefont {Wilczek},\ and\ \citenamefont
  {Borish}}]{cotler2016entanglement}%
  \BibitemOpen
  \bibfield  {author} {\bibinfo {author} {\bibfnamefont {J.}~\bibnamefont
  {Cotler}}, \bibinfo {author} {\bibfnamefont {F.}~\bibnamefont {Wilczek}}, \
  and\ \bibinfo {author} {\bibfnamefont {V.}~\bibnamefont {Borish}},\
  }\href@noop {} {\bibfield  {journal} {\bibinfo  {journal} {arXiv preprint
  arXiv:1607.05719}\ } (\bibinfo {year} {2016})}\BibitemShut {NoStop}%
\bibitem [{\citenamefont {Qu}\ \emph {et~al.}(2019)\citenamefont {Qu},
  \citenamefont {Cotler}, \citenamefont {Ma}, \citenamefont {Guan},
  \citenamefont {Zheng}, \citenamefont {Xie}, \citenamefont {Chen},
  \citenamefont {Zhang}, \citenamefont {Wilczek},\ and\ \citenamefont
  {Pan}}]{qu2019color}%
  \BibitemOpen
  \bibfield  {author} {\bibinfo {author} {\bibfnamefont {L.-Y.}\ \bibnamefont
  {Qu}}, \bibinfo {author} {\bibfnamefont {J.}~\bibnamefont {Cotler}}, \bibinfo
  {author} {\bibfnamefont {F.}~\bibnamefont {Ma}}, \bibinfo {author}
  {\bibfnamefont {J.-Y.}\ \bibnamefont {Guan}}, \bibinfo {author}
  {\bibfnamefont {M.-Y.}\ \bibnamefont {Zheng}}, \bibinfo {author}
  {\bibfnamefont {X.}~\bibnamefont {Xie}}, \bibinfo {author} {\bibfnamefont
  {Y.-A.}\ \bibnamefont {Chen}}, \bibinfo {author} {\bibfnamefont
  {Q.}~\bibnamefont {Zhang}}, \bibinfo {author} {\bibfnamefont
  {F.}~\bibnamefont {Wilczek}}, \ and\ \bibinfo {author} {\bibfnamefont
  {J.-W.}\ \bibnamefont {Pan}},\ }\href@noop {} {\bibfield  {journal} {\bibinfo
   {journal} {Physical Review Letters}\ }\textbf {\bibinfo {volume} {123}},\
  \bibinfo {pages} {243601} (\bibinfo {year} {2019})}\BibitemShut {NoStop}%
\bibitem [{\citenamefont {Qu}\ \emph {et~al.}(2020)\citenamefont {Qu},
  \citenamefont {Liu}, \citenamefont {Cotler}, \citenamefont {Ma},
  \citenamefont {Guan}, \citenamefont {Zheng}, \citenamefont {Yao},
  \citenamefont {Xie}, \citenamefont {Chen}, \citenamefont {Zhang} \emph
  {et~al.}}]{qu2020chromatic}%
  \BibitemOpen
  \bibfield  {author} {\bibinfo {author} {\bibfnamefont {L.-Y.}\ \bibnamefont
  {Qu}}, \bibinfo {author} {\bibfnamefont {L.-C.}\ \bibnamefont {Liu}},
  \bibinfo {author} {\bibfnamefont {J.}~\bibnamefont {Cotler}}, \bibinfo
  {author} {\bibfnamefont {F.}~\bibnamefont {Ma}}, \bibinfo {author}
  {\bibfnamefont {J.-Y.}\ \bibnamefont {Guan}}, \bibinfo {author}
  {\bibfnamefont {M.-Y.}\ \bibnamefont {Zheng}}, \bibinfo {author}
  {\bibfnamefont {Q.}~\bibnamefont {Yao}}, \bibinfo {author} {\bibfnamefont
  {X.}~\bibnamefont {Xie}}, \bibinfo {author} {\bibfnamefont {Y.-A.}\
  \bibnamefont {Chen}}, \bibinfo {author} {\bibfnamefont {Q.}~\bibnamefont
  {Zhang}},  \emph {et~al.},\ }\href@noop {} {\bibfield  {journal} {\bibinfo
  {journal} {Optics Express}\ }\textbf {\bibinfo {volume} {28}},\ \bibinfo
  {pages} {32294} (\bibinfo {year} {2020})}\BibitemShut {NoStop}%
\bibitem [{\citenamefont {Takesue}(2008)}]{takesue2008erasing}%
  \BibitemOpen
  \bibfield  {author} {\bibinfo {author} {\bibfnamefont {H.}~\bibnamefont
  {Takesue}},\ }\href@noop {} {\bibfield  {journal} {\bibinfo  {journal}
  {Physical review letters}\ }\textbf {\bibinfo {volume} {101}},\ \bibinfo
  {pages} {173901} (\bibinfo {year} {2008})}\BibitemShut {NoStop}%
\bibitem [{\citenamefont {Raymer}\ \emph {et~al.}(2010)\citenamefont {Raymer},
  \citenamefont {Van~Enk}, \citenamefont {McKinstrie},\ and\ \citenamefont
  {McGuinness}}]{raymer2010interference}%
  \BibitemOpen
  \bibfield  {author} {\bibinfo {author} {\bibfnamefont {M.}~\bibnamefont
  {Raymer}}, \bibinfo {author} {\bibfnamefont {S.}~\bibnamefont {Van~Enk}},
  \bibinfo {author} {\bibfnamefont {C.}~\bibnamefont {McKinstrie}}, \ and\
  \bibinfo {author} {\bibfnamefont {H.}~\bibnamefont {McGuinness}},\
  }\href@noop {} {\bibfield  {journal} {\bibinfo  {journal} {Optics
  Communications}\ }\textbf {\bibinfo {volume} {283}},\ \bibinfo {pages} {747}
  (\bibinfo {year} {2010})}\BibitemShut {NoStop}%
\bibitem [{\citenamefont {De~Greve}\ \emph {et~al.}(2012)\citenamefont
  {De~Greve}, \citenamefont {Yu}, \citenamefont {McMahon}, \citenamefont
  {Pelc}, \citenamefont {Natarajan}, \citenamefont {Kim}, \citenamefont {Abe},
  \citenamefont {Maier}, \citenamefont {Schneider}, \citenamefont {Kamp},\ and\
  \citenamefont {Höfling}}]{de2012quantum}%
  \BibitemOpen
  \bibfield  {author} {\bibinfo {author} {\bibfnamefont {K.}~\bibnamefont
  {De~Greve}}, \bibinfo {author} {\bibfnamefont {L.}~\bibnamefont {Yu}},
  \bibinfo {author} {\bibfnamefont {P.~L.}\ \bibnamefont {McMahon}}, \bibinfo
  {author} {\bibfnamefont {J.~S.}\ \bibnamefont {Pelc}}, \bibinfo {author}
  {\bibfnamefont {C.~M.}\ \bibnamefont {Natarajan}}, \bibinfo {author}
  {\bibfnamefont {N.~Y.}\ \bibnamefont {Kim}}, \bibinfo {author} {\bibfnamefont
  {E.}~\bibnamefont {Abe}}, \bibinfo {author} {\bibfnamefont {S.}~\bibnamefont
  {Maier}}, \bibinfo {author} {\bibfnamefont {C.}~\bibnamefont {Schneider}},
  \bibinfo {author} {\bibfnamefont {M.}~\bibnamefont {Kamp}}, \ and\ \bibinfo
  {author} {\bibfnamefont {S.}~\bibnamefont {Höfling}},\ }\href@noop {}
  {\bibfield  {journal} {\bibinfo  {journal} {Nature}\ }\textbf {\bibinfo
  {volume} {491}},\ \bibinfo {pages} {421} (\bibinfo {year}
  {2012})}\BibitemShut {NoStop}%
\bibitem [{\citenamefont {Twiss}\ and\ \citenamefont
  {Brown}(1957)}]{twiss1957question}%
  \BibitemOpen
  \bibfield  {author} {\bibinfo {author} {\bibfnamefont {R.}~\bibnamefont
  {Twiss}}\ and\ \bibinfo {author} {\bibfnamefont {R.~H.}\ \bibnamefont
  {Brown}},\ }\href@noop {} {\bibfield  {journal} {\bibinfo  {journal}
  {Nature}\ }\textbf {\bibinfo {volume} {179}},\ \bibinfo {pages} {1128}
  (\bibinfo {year} {1957})}\BibitemShut {NoStop}%
\bibitem [{\citenamefont {Baym}(1998{\natexlab{b}})}]{baym1998physics}%
  \BibitemOpen
  \bibfield  {author} {\bibinfo {author} {\bibfnamefont {G.}~\bibnamefont
  {Baym}},\ }\href@noop {} {\bibfield  {journal} {\bibinfo  {journal} {Acta
  Physica Polonica. Series B}\ }\textbf {\bibinfo {volume} {29}},\ \bibinfo
  {pages} {1839} (\bibinfo {year} {1998}{\natexlab{b}})}\BibitemShut {NoStop}%
\bibitem [{\citenamefont {van Cittert}(1934)}]{van1934wahrscheinliche}%
  \BibitemOpen
  \bibfield  {author} {\bibinfo {author} {\bibfnamefont {P.~H.}\ \bibnamefont
  {van Cittert}},\ }\href@noop {} {\bibfield  {journal} {\bibinfo  {journal}
  {Physica}\ }\textbf {\bibinfo {volume} {1}},\ \bibinfo {pages} {201}
  (\bibinfo {year} {1934})}\BibitemShut {NoStop}%
\bibitem [{\citenamefont {Zernike}(1938)}]{zernike1938concept}%
  \BibitemOpen
  \bibfield  {author} {\bibinfo {author} {\bibfnamefont {F.}~\bibnamefont
  {Zernike}},\ }\href@noop {} {\bibfield  {journal} {\bibinfo  {journal}
  {Physica}\ }\textbf {\bibinfo {volume} {5}},\ \bibinfo {pages} {785}
  (\bibinfo {year} {1938})}\BibitemShut {NoStop}%
\bibitem [{\citenamefont {Nunez}\ \emph {et~al.}(2012)\citenamefont {Nunez},
  \citenamefont {Holmes}, \citenamefont {Kieda},\ and\ \citenamefont
  {LeBohec}}]{nunez2012high}%
  \BibitemOpen
  \bibfield  {author} {\bibinfo {author} {\bibfnamefont {P.~D.}\ \bibnamefont
  {Nunez}}, \bibinfo {author} {\bibfnamefont {R.}~\bibnamefont {Holmes}},
  \bibinfo {author} {\bibfnamefont {D.}~\bibnamefont {Kieda}}, \ and\ \bibinfo
  {author} {\bibfnamefont {S.}~\bibnamefont {LeBohec}},\ }\href@noop {}
  {\bibfield  {journal} {\bibinfo  {journal} {Monthly Notices of the Royal
  Astronomical Society}\ }\textbf {\bibinfo {volume} {419}},\ \bibinfo {pages}
  {172} (\bibinfo {year} {2012})}\BibitemShut {NoStop}%
\bibitem [{\citenamefont {Kieda}\ and\ \citenamefont
  {Matthews}(2017)}]{kieda2017stellar}%
  \BibitemOpen
  \bibfield  {author} {\bibinfo {author} {\bibfnamefont {D.}~\bibnamefont
  {Kieda}}\ and\ \bibinfo {author} {\bibfnamefont {N.}~\bibnamefont
  {Matthews}},\ }\href@noop {} {\bibfield  {journal} {\bibinfo  {journal}
  {arXiv preprint arXiv:1709.03956}\ } (\bibinfo {year} {2017})}\BibitemShut
  {NoStop}%
\bibitem [{\citenamefont {Abeysekara}\ \emph {et~al.}(2020)\citenamefont
  {Abeysekara}, \citenamefont {Benbow}, \citenamefont {Brill}, \citenamefont
  {Buckley}, \citenamefont {Christiansen}, \citenamefont {Chromey},
  \citenamefont {Daniel}, \citenamefont {Davis}, \citenamefont {Falcone},
  \citenamefont {Feng} \emph {et~al.}}]{abeysekara2020demonstration}%
  \BibitemOpen
  \bibfield  {author} {\bibinfo {author} {\bibfnamefont {A.}~\bibnamefont
  {Abeysekara}}, \bibinfo {author} {\bibfnamefont {W.}~\bibnamefont {Benbow}},
  \bibinfo {author} {\bibfnamefont {A.}~\bibnamefont {Brill}}, \bibinfo
  {author} {\bibfnamefont {J.}~\bibnamefont {Buckley}}, \bibinfo {author}
  {\bibfnamefont {J.}~\bibnamefont {Christiansen}}, \bibinfo {author}
  {\bibfnamefont {A.}~\bibnamefont {Chromey}}, \bibinfo {author} {\bibfnamefont
  {M.}~\bibnamefont {Daniel}}, \bibinfo {author} {\bibfnamefont
  {J.}~\bibnamefont {Davis}}, \bibinfo {author} {\bibfnamefont
  {A.}~\bibnamefont {Falcone}}, \bibinfo {author} {\bibfnamefont
  {Q.}~\bibnamefont {Feng}},  \emph {et~al.},\ }\href@noop {} {\bibfield
  {journal} {\bibinfo  {journal} {Nature Astronomy}\ ,\ \bibinfo {pages} {1}}
  (\bibinfo {year} {2020})}\BibitemShut {NoStop}%
\end{thebibliography}%

\end{document}